\documentclass{aa}
\usepackage{graphicx}
\def\ltapprox{\raise 2pt \hbox {$<$} \kern-1.1em \lower 5pt \hbox {$\approx$}}
\def\ltsim{\raise 2pt \hbox {$<$} \kern-1.1em \lower 4pt \hbox {$\sim$}}
\def\gtsim{\raise 2pt \hbox {$>$} \kern-1.1em \lower 4pt \hbox {$\sim$}}
\def\arcsec{$^{\prime\prime}\,$}

\usepackage{txfonts}

\begin{document}

\title{Multifrequency VLA radio observations of the X-ray cavity cluster
of galaxies RBS797: evidence of differently oriented jets}

\author{M. Gitti\inst{1}
\and L. Feretti\inst{2}
\and S. Schindler\inst{1}
}

\institute{Institut f\"ur Astrophysik, Leopold-Franzens Universit\"at
Innsbruck, Technikerstra\ss e 25, A-6020 Innsbruck, Austria
\and
Istituto di Radioastronomia - INAF
via Gobetti 101, I-40129 Bologna, Italy
}

\offprints{Myriam Gitti,
\email{myriam.gitti@uibk.ac.at}}

\authorrunning{M. Gitti}
\titlerunning{Multifrequency VLA observations of the galaxy cluster RBS797}
%\date{5 August 2005 - Submitted - Please do not circulate}
\date{Received / Accepted}

\abstract{ We report on the peculiar activity of the radio source
located at the center of the cooling flow cluster RBS797 (z=0.35),
the first distant cluster in which two pronounced X-ray cavities
have been discovered. New multifrequency (1.4, 4.8, and 8.4 GHz)
observations obtained with the Very Large Array clearly reveal the
presence of radio emission on three different scales showing
orientation in different directions, all of which indicates that
RBS797 represents a very peculiar case. The lowest resolution
images show large-scale radio emission characterized by amorphous
morphology and a steep spectrum, extended on a scale of hundreds
of kpc. On a scale of tens of kpc, there is evidence of 1.4 GHz
radio emission elongated in the northeast-southwest direction
exactly towards the holes detected in X-rays. The highest
resolution image shows the details of the innermost 4.8 GHz radio
jets on a kpc scale; they are remarkably oriented in a direction
that is perpendicular to that of the extended structure detected
at a lower resolution. We therefore find evidence of a strong
interaction between the central radio source and the intra-cluster
medium in RBS797. We suggest a scenario in which the 1.4 GHz
emission filling the X-ray cavities consists of buoyant bubbles of
radio emitting plasma that are created by twin jets in the past
and whose expansion has displaced the thermal gas that was
formerly in the X-ray holes, whereas the two jets visible at 4.8
GHz are related to the present nuclear activity that has restarted
at a different position angle from the original outburst that
created the outer radio lobes. The total radio luminosity is $\sim
10^{42} {\rm erg ~ s}^{-1}$, corresponding to a factor of a few
thousand times less than the estimated cooling luminosity.
 }
\maketitle

%%%%%%%%%%%%%%%%%%%%%%%%%%%%%%%%%%%%%%%%%%%%%%%%%%%%%%%%%%%%%%%%%%%%%%%%%%%%%%

\section{Introduction}
\label{intro}

An efficient cooling of the intra-cluster medium (ICM) is expected
to occur in the clusters of galaxies with a sufficiently high gas
density for the radiative cooling time due to X-ray emission to be
shorter than the Hubble time (or the time since its last major
merger). In the absence of any balancing heating mechanisms, the
gas in the central regions of these clusters is expected to cool
down and flow slowly inwards in order to maintain hydrostatic
equilibrium (for a review of the standard \textit{cooling flow}
model, see Fabian 1994).
Recent X-ray observations with {\it Chandra} and {\it XMM-Newton},
albeit confirming the existence of inwardly decreasing temperature
gradients and short central cooling times, have altered this
simple picture of steady cooling flows and show that little of the
gas cools below $\sim$ 1-2 keV (e.g. David et al. 2001, Johnston
et al. 2002, Peterson et al. 2003 and references therein). The
lack of observations of cooler gas represents an open question
that is often referred to as the so called {\it 'cooling flow
problem'}.

Besides the hypothesis that the normal signatures of cooling below
1-2 keV are somehow suppressed \footnote{Different possibilities
have been investigated in this context, including: absorption
(Peterson et al. 2001, Fabian et al. 2001), inhomogeneous
metallicity (Fabian et al. 2001, Morris \& Fabian 2003), the
emerging of the missing X--ray luminosity in other bands, like
ultraviolet, optical and infrared due to mixing with cooler
gas/dust (Fabian et al. 2001, 2002a, Mathews \& Brighenti 2003),
and radio due to particle re--acceleration (Gitti et al. 2002,
2004).}, in order to avoid the expected rapid cooling to very low
temperature, there must be some form of energy input into the ICM
to balance the cooling.
Currently the best candidates for supplying the energy are
cluster-center radio sources that provide heating through
processes associated with relativistic AGN outflows (e.g., Rosner
\& Tucker 1989; Tabor \& Binney 1993; Churazov et al. 2001;
Br{\"u}ggen \& Kaiser 2001; Kaiser \& Binney 2002; Ruszkowski \&
Begelman 2002; Brighenti \& Mathews 2003). Other proposed heating
mechanisms include electron thermal conduction from the outer
regions of clusters (Tucker \& Rosner 1983; Voigt et al. 2002;
Fabian et al. 2002b; Zakamska \& Narayan 2003), continuous
subcluster merging (Markevitch et al. 2001), contribution of the
gravitational potential of the cluster core (Fabian 2003),
feedback from intra-cluster supernovae (Domainko et al. 2004),
etc.

As shown by Burns (1990), the vast majority of cooling flow
clusters contain powerful radio sources associated with the
central cD galaxies. These radio sources have a profound effect on
the ICM, as the radio lobes displace the X-ray emitting gas,
thereby creating X-ray deficient ``holes'' or ``bubbles''. High
resolution X-ray images of cooling flows in galaxy clusters taken
with \textit{Chandra} indicate that indeed the central hot gas in
these systems is not smoothly distributed but is cavitated on
scales ranging from a few to a few tens of kpc, often
approximately coincident with lobes of extended radio emission
(e.g., Hydra A: McNamara et al. 2000, David et al. 2001, Allen et
al. 2001; Perseus: Churazov et al. 2000, B\"ohringer et al. 1993,
Fabian et al. 2000; A2052: Blanton et al. 2001, 2003; A2597:
McNamara et al. 2001; RBS797: Schindler et al. 2001, De Filippis
et al. 2002; A496: Dupke \& White 2002; MKW 3s: Mazzotta et al.
2002; A2199: Johnstone et al. 2002; A4059: Heinz et al. 2002;
Virgo: Young et al. 2002; Centaurus: Sanders \& Fabian 2002;
Cygnus A: Smith et al. 2002; A262: Blanton et al. 2004; A2597:
Pollack et al. 2005, Clarke et al. 2005 ). Giant cavities, on
scales of hundreds of kpc, have recently been found in the
optically poor cluster MS0735.6+7241 (McNamara et al. 2005) and in
Hercules A (Nulsen et al. 2005).

Comparisons of the total power emitted from the central radio source
with the luminosity of cooling gas have shown that the energy emitted from
a central radio source is sufficient to offset the cooling of the ICM
in the centers of clusters, at least in some cases
(e.g.,
Hydra A: David et al. 2001;
A2052: Blanton et al. 2003).
By performing a systematic study of radio-induced X-ray cavities,
B\^{\i}rzan et al (2004) show that the energy input from the
central radio source is sufficient to balance cooling in roughly
half of the 16 systems they study, at least for short time
periods. It is interesting to note that the objects balancing the
central cooling show a large spread in both radio luminosity and
spectral index.
The details of the transportation of the radio source energy to the ICM are
still somehow obscure; however, the scenario of radio source heating
remains a promising solution for the problem of the missing cool gas in
the cooling flow model.

The X-ray luminous, distant galaxy cluster RBS797 (z=0.35) was
discovered in the ROSAT All-Sky Survey (Schwope et al. 2000).
Observations with \textit{Chandra} reveal two pronounced X-ray
minima that are located opposite to each other with respect to the
cluster center (Schindler et al. 2001). In the NRAO VLA Sky Survey
(NVSS), an unresolved radio source of 20 mJy is present at the
center of RBS797, and a spectrum of the central cluster galaxy
shows emission lines; hence, the radio emission is likely to
originate from an AGN. Preliminary results from low resolution VLA
radio observations confirm the presence of a strong radio source
positioned in the center of the cluster (De Filippis et al. 2002).
As in the case of the X-ray holes observed in other clusters, the
X-ray depressions detected in RBS797 suggest an interaction
between the central radio galaxy and the ICM. This is a special
case, though, since RBS797 has given us the chance to observe
these features for the first time in a relatively distant cluster;
and furthermore, the minima in this cluster are very symmetric and
very deep compared to similar features found in other clusters.

Studying the interaction between the central radio galaxies of
clusters and the ICM is of general interest, as such interaction
can have numerous effects on the ICM. For example, the energy
input from active galaxies into the ICM can be one of the reasons
why poor clusters do not follow the general X-ray luminosity -
temperature relation (Ponman et al. 1999). The additional
non-gravitational heating required could be partly supplied by
active galaxies. Furthermore, relativistic particles are injected
into the ICM by active galaxies, and although they lose their
energy to the ICM relatively fast, they might be re-accelerated
later, probably in shock waves or turbulence generated by cluster
mergers (e.g., Tribble 1993, Brunetti et al. 2001, En{\ss}lin et
al. 1998) or turbulence in the cooling flows (Gitti et al. 2002).
These re-accelerated particles are very likely the source for the
radio halos, relics, and mini-halos found in a number of galaxy
clusters (for a more detailed discussion about diffuse radio
emission observed in clusters see e.g., Feretti et al. 2004 and
Kempner et al. 2004), and probably also for the non-thermal hard
X-ray excess observed in some galaxy clusters (Fusco-Femiano 1999,
2001). Finally, the material transported from the galaxy into the
ICM is probably metal-enriched, so that this process could
contribute to the metal enrichment of the ICM. Typically, the
metallicity of the ICM is about 0.3 solar; i.e. a lot of heavy
elements must have been ejected by the cluster galaxies. Up to now
it is not clear which of the suggested processes (galactic winds,
ram-pressure stripping, galaxy-galaxy interaction, or ejection
from active galaxies) is dominating at different redshifts (e.g.,
Schindler et al. 2005).

In this paper we present the VLA observations of the radio source in RBS797
and perform a detailed comparison of its structure with the structure of the
ICM derived from earlier \textit{Chandra} X-ray observations.
In particular, we focus on the interaction between the radio source and the
X-ray emitting ICM in order to get a
comprehensive picture of the physical processes in clusters of galaxies.
RSB797, whose central cluster galaxy is located at RA (J2000):
09$^{\rm h}$ 47$^{\rm m}$ 12$^{\rm s}$.5, Dec (J2000):
76$^{\circ}$ 23$'$ 14$''$.0, is at a redshift $ z= 0.35$. With
$H_0 = 70 \mbox{ km s}^{-1} \mbox{ Mpc}^{-1}$, and
$\Omega_M=1-\Omega_{\Lambda}=0.3$, the luminosity distance is 1858
Mpc and 1 arcsec corresponds to 4.8 kpc. The radio spectral index
$\alpha$ is defined as $S_{\nu} \propto \nu^{\alpha}$.

%%%%%%%%%%%%%%%%%%%%%%%%%%%%%%%%%%%%%%%%%%%%%%%%%%%%%%%%%%%%%%%%%%%%%%%%%%%%%%

\section{Observations and data reduction}
\label{observations}

\begin{table}
\begin{center}
\caption{VLA data archive} \hskip -.2truein
\begin{tabular}{cccccc}
\hline
\hline
Observation & Frequency & Bandwidth & Array & Time\\
Date & (MHz) & (MHz) & ~ & (h) \\
\hline
~&~&~&~\\
Sep-2001 & 8435/8485 & 50.0 & D & 5.0\\
Mar-2002 & 1435/1515 & 50.0 & A & 4.5\\
Jul-2002 & 1435/1515 & 50.0 & B & 4.8\\
Dec-2004 & 4835/4885 & 50.0 & A & 7.0\\
\hline
\label{vladata.tab}
\end{tabular}
\end{center}
\end{table}

Very Large Array\footnote{The Very Large Array (VLA) is a facility
of the National Radio Astronomy Observatory (NRAO). The NRAO is a
facility of the National Science Foundation, operated under
cooperative agreement by Associated Universities, Inc.}
observations of the radio source RBS797 were made at three
different frequencies (1.4 GHz, 4.8 GHz, 8.4 GHz) in different
configurations (see Table \ref{vladata.tab} for details). In all
observations the source 3C 286 was used as the primary flux
density calibrator, while the sources 1044+809 and 0713+438 were
used as secondary phase and polarization calibrators,
respectively.

Data reduction was done using the NRAO AIPS (Astronomical Image
Processing System) package. Accurate editing of the uv data was
applied to identify and remove bad data. At 1.4 GHz, only the
channel centered at 1435 MHz was used because of strong
interferences in the other channel. The 1.4 GHz data from the two
different configurations were reduced separately, in order to
analyze the possible existence of spurious features. Images at
different resolutions were also obtained by adding the data
together from the two configurations and by specifying appropriate
values of the parameters  UVTAPER and ROBUST in the AIPS task
IMAGR. Images were produced by following the standard procedures:
Calibration, Fourier-Transform, Clean and Restore.
Self-calibration was applied to remove residual phase variations.
Images in the Stokes parameters I, Q, and U were produced using
the AIPS task IMAGR. The images of the polarized intensity, the
fractional polarization, and the position angle of polarization
were derived from the I, Q, and U images. The final images (Figs.
\ref{mappa-8.4-D}, \ref{zoom}a, \ref{zoom}b, and \ref{zoom}c) show
the contours of the total intensity.
%overlaid on the polarization vectors.
%The vector orientation shows the projected E-field and their length is
%proportional to the fractional polarization.
%In the figures, only vectors with an error in the fractional polarization
%less than 10\% were considered.

%%%%%%%%%%%%%%%%%%%%%%%%%%%%%%%%%%%%%%%%%%%%%%%%%%%%%%%%%%%%%%%%%%%%%%%%%%%%%

\section{Results}
\label{results}

\begin{figure}[ht]
\includegraphics{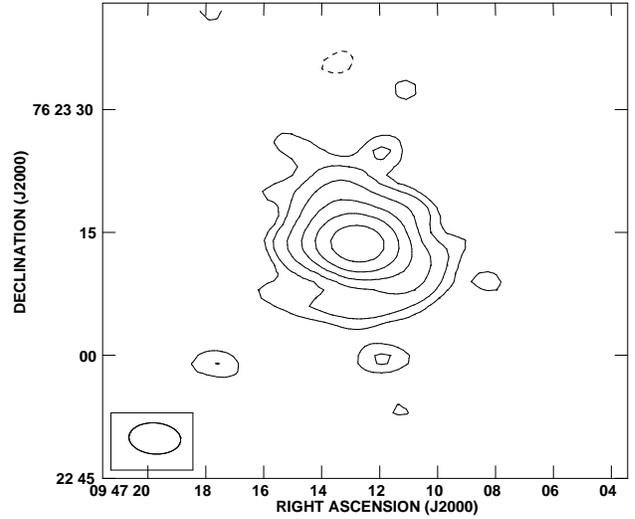} \vspace{6.8cm} \caption{ 8.4 GHz VLA
map of RBS797 at a resolution of $6''.3 \times 3''.8$ (the beam is
shown in the lower lefthand corner). The contour levels are
$-0.03$ (dashed), 0.03, 0.05, 0.10, 0.20, 0.40, 0.85, 1.70, 3.50
mJy/beam. The r.m.s. noise is 0.01 mJy/beam. } \label{mappa-8.4-D}
\end{figure}

Figure \ref{mappa-8.4-D} shows the radio map of RBS797 observed at
8.4 GHz with the VLA in D configuration, with a restoring beam of
$6''.3 \times 3''.8$. The source shows an amorphous morphology
elongated in the NE-SW direction with a total size of
$\sim$25\arcsec (120 kpc). The source has a total flux density of
$\simeq 3.02 \pm 0.03$ mJy, with a contribution of $\simeq 1.44
\pm 0.02$ mJy coming from the central $\sim$15\arcsec region. The
source appears polarized at levels of $\simeq 10-20$\% in a region
extending $\sim7$\arcsec$\times5$\arcsec (34 $\times$ 24 kpc ) out
from the center.

\begin{figure}[t]
\includegraphics{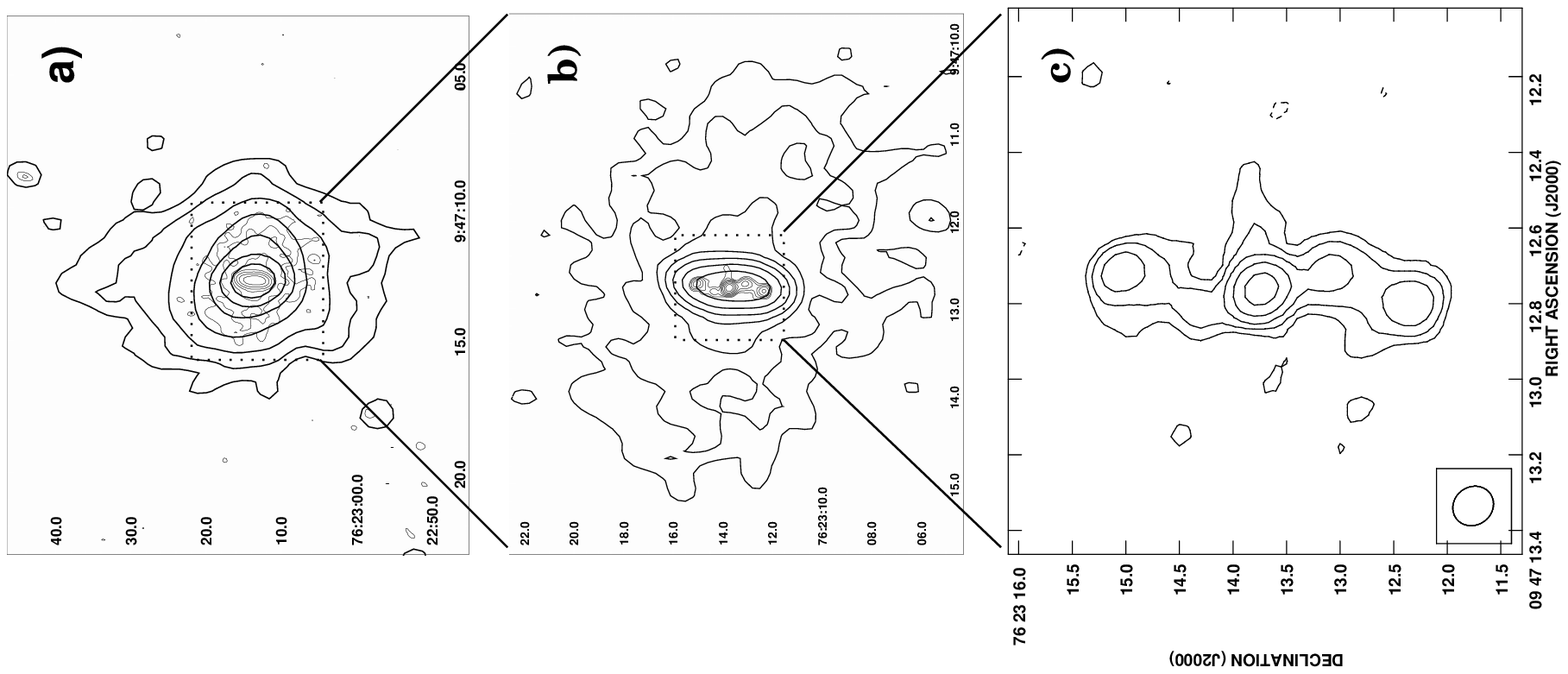}
\vspace{21.3cm}
\caption{
}
\label{zoom}
\end{figure}

Figure \ref{zoom}a shows the radio map of RBS797 observed at 1.4
GHz with the VLA in B configuration, with a restoring beam of
$4''.9 \times 3''.8$. At this lower frequency and low resolution,
the source shows its largest extent, with diffuse emission
extended by approximately 45\arcsec (216 kpc) in the N-S direction
and $\sim$30\arcsec (144 kpc) in the other direction. The
morphology is quite regular, although slightly elongated to the
north. The source has a total flux density of $\simeq 26.0 \pm
0.3$ mJy, with a contribution of $\simeq 11.5 \pm 0.1$ mJy coming
from the central region. The source appears polarized at levels of
$\simeq 10-20$\% in a region extending
$\sim10$\arcsec$\times15$\arcsec out from the center.

Figure \ref{zoom}b shows the radio map of RBS797 observed at 1.4
GHz with the VLA in A configuration with a restoring beam of
$1''.5 \times 1''.1$. The higher resolution allows us to detect
details of the central region. The source has a structure
elongated in the NE-SW direction, as in the 8.4 GHz map presented
in Fig. \ref{mappa-8.4-D}, although with a smaller total extent
(18\arcsec $\sim$ 86 kpc). There is also a hint of possible faint
jets emanating from the central component. The bright inner radio
emission is resolved, showing an elongation along the N-S
direction. The source has a total flux density of $\simeq 17.9 \pm
0.2$ mJy, with a contribution of $\simeq 12.1 \pm 0.1$ mJy coming
from the central region (inner jets included, while the core alone
has a flux density of $\simeq 10.8 \pm 0.1$ mJy). The central
region appears polarized at levels of $\simeq 10-15$\%. Note that
the structure shown here corresponds to the central
$\sim$15\arcsec (72 kpc) region in Fig. \ref{zoom}a, which has a
total flux density of $\simeq 18.0 \pm 0.2$ mJy in agreement with
these results at higher resolution.

%\begin{figure}[ht]
%\special{
%psfile=figure/new/RBS797-1.4-M.ps hoffset=-10 voffset=-275 vscale=45
%hscale=45 angle=0
%}
%\vspace{6.5cm}
%\caption{
%\textbf{1.4 GHz VLA map of RBS797 at a resolution of $3''.1 \times 2''.7$
%\textbf{(the beam is shown in the lower-left corner)}.
%The contour levels are $-0.05$ (dashed), 0.05, 0.10, 0.20, 0.40, 0.80, 1.50,
%3.00 mJy/beam. The r.m.s. noise is 0.02 mJy/beam.}
%}
%\label{mappa-1.4-comb}
%\end{figure}

The radio images  presented in Figs. \ref{zoom}a and \ref{zoom}b
have been obtained with the data from single configurations. We
also obtained radio images at low ($\sim 5''$) and high ($\sim
1''$) resolution by adding together the data from the two
configurations and by specifying appropriate values of the
parameters in the AIPS task IMAGR (UVTAPER=50, ROBUST=5 and
UVTAPER=0, ROBUST=-5 for low and high resolution, respectively).
The results are consistent with those obtained from single
configurations. An image with the resolution of $\sim 3''$ was
obtaiend from the combination of the data from the two
configurations by setting UVTAPER=60. This image (not shown here)
is fully consistent with the other images presented in the paper.

Figure \ref{zoom}c shows the radio map of RBS797 observed at 4.8
GHz with the VLA in A configuration with a restoring beam of
$0''.39 \times 0''.35$. The high resolution allows us to determine
the position of the nucleus accurately, which is located at RA
(J2000): 09$^{\rm h}$ 47$^{\rm m}$ 12$^{\rm s}$.76, Dec (J2000):
76$^{\circ}$ 23$'$ 13$''$.74.

\newpage

~\\
~\\
\newpage
\vspace{3.cm}
{\small
~\\
\line(1,0){90}
\textbf{Caption of Fig. 2}
\line(1,0){90}
\\
Observations at different resolutions show orientations in
different directions of the radio structures. The highest
resolution image on kpc scale shows the details for the innermost
4.8 GHz radio jets, which clearly point to the N-S direction
(panel c). Remarkably, the inner jets are oriented in a
direction almost perpendicular to that of the extended structure
detected at lower resolution, which is elongated in the NE-SW
direction (panel b). The lowest resolution images show large-scale
radio emission with amorphous morphology, slightly extended in the
N-S direction (see panel a). See text (Sect.
\ref{discussion}) for more details.
\\
\textbf{a)}
Black contours: 1.4 GHz VLA map of RBS797 at a resolution
of $4''.9 \times 3''.8$.
The contour levels are $-0.08$ (dashed), 0.08, 0.15, 0.30, 0.60, 1.20, 2.50
mJy/beam. The r.m.s. noise is 0.03 mJy/beam.
Overlayed in grey are the 1.4 GHz contour plot at $\sim 1$\arcsec resolution
(the beam and r.m.s. noise are the same as in Fig. \ref{zoom}b).
\\
\textbf{b)}
Black contours: 1.4 GHz VLA map of RBS797 at a resolution
of $1''.5 \times 1''.1$. The contour levels are $-0.05$ (dashed), 0.05, 0.10,
0.20, 0.40, 0.80, 1.50, 3.00 mJy/beam. The r.m.s. noise is 0.02 mJy/beam.
Overlayed in grey are the 4.8 GHz contour plot at $\sim$0\arcsec.4 resolution
(the beam and r.m.s. noise are the same as in Fig. \ref{zoom}c).
\\
\textbf{c)}
4.8 GHz VLA map of RBS797 at a resolution of $0''.39 \times 0''.35$
(the beam is shown in the lower-left corner).
The contour levels are $-0.04$ (dashed), 0.04, 0.08, 0.16, 0.32, 0.64, 1.28,
2.50 mJy/beam. The r.m.s. noise is 0.01 mJy/beam.
~\\
\line(1,0){80}
\textbf{End of caption of Fig. 2}
\line(1,0){80}
}
~\\

The N-S elongation structure of the bright inner radio emission,
detected in Fig. \ref{zoom}b, is fully unveiled here: the source
clearly shows two jets pointing in the N-S direction and extending
out from the core to a distance of approximately 2.8\arcsec (13.4
kpc). Two spots of enhanced brightness are visible at the outer
edges of the jets. A feature extending to the west out to a
distance of $\sim$1.9\arcsec (9 kpc) from the core is also visible
($\sim 5 \sigma$ detection). The source has a total flux density
of $\simeq 2.63 \pm 0.03$ mJy, with a contribution of $\simeq 0.83
\pm 0.02$ mJy coming from the nuclear region. The central region
and the two brighter spots appear strongly polarized at levels of
$\simeq 10-35$\%.

Very remarkable is that the two jets visible in Fig. \ref{zoom}c
are oriented in a direction perpendicular to that of the extended
structures detected at lower resolution (Figs. \ref{mappa-8.4-D}
and \ref{zoom}b). %%
In order to show the relation of the diffuse structure to the
finer scale structure detected at higher resolution, we display in
Fig. \ref{zoom}a the overlay of the 1.4 GHz map at $\sim 4$\arcsec
resolution onto the 1.4 GHz map at $\sim 1$\arcsec resolution, and
then in Fig. \ref{zoom}b the overlay of the same 1.4 GHz map at
arcsec resolution onto the 4.8 GHz map at subarcsec resolution.

The radio results are summarized in Table \ref{risradio.tab}, where we also
report the monochromatic radio power at each frequency calculated as
\begin{equation}
P_{\nu}=4 \pi \, D_{\rm L}^2 \, S_{\nu} \, (1+z)^{-(\alpha +1)}
\label{radio_power.eq}
\end{equation}
where $D_{\rm L}$ is the luminosity distance, $S_{\nu}$ the flux
density in Janskys\footnote{1 Jy = $10^{-26} \, {\rm W} \, {\rm
Hz}^{-1} {\rm m}^{-2}$} at the frequency $\nu$, and
$(1+z)^{-(\alpha +1)}$ the \textit{K-correction} term (Petrosian
\& Dickey 1973), which in our case is negligible ($\alpha \sim
-1$).

%%%%%%%%%%%%%%%%%%%%%%%%%%%%%%%%%%%%%%%%%%%%%%%%%%%%%%%%%%%%%%%%%%%%%%%%%%%%%%

\subsection{Spectral index}

The dependence of the source's appearance on frequency and
resolution reflects both the diffuse nature of the extended
emission and the steep, but position-dependent, spectrum of the
radio emission. By comparing the peak flux densities at 8.4 GHz in
array D, 1.4 GHz in array B, and the total flux density at 4.8 GHz
in array A, we estimated that the central region has a spectral
index of $\alpha^{8.4}_{4.8} = -0.98 \pm 0.03$,
$\alpha^{4.8}_{1.4} = -1.09 \pm 0.01$, and $\alpha^{8.4}_{1.4} =
-1.06 \pm 0.01$. We note that these spectral values, which refer
to the emission visible in Fig. \ref{zoom}a, are steeper than
those commonly expected in nuclear-jet structures.

\begin{figure}[ht]
\includegraphics{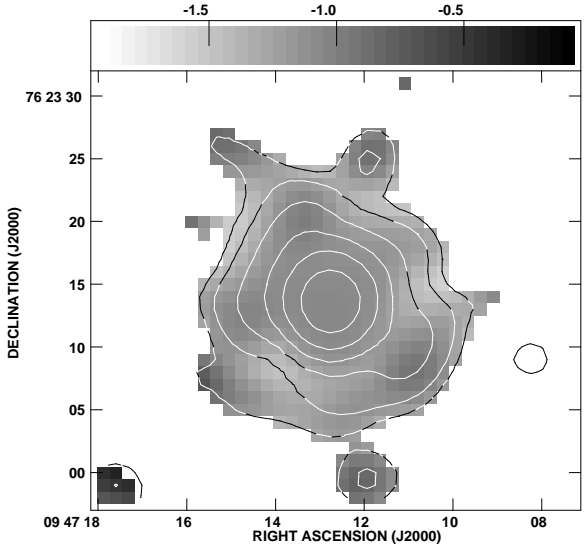} \vspace{6.8cm} \caption{ Spectral
index distribution between $\nu=8.4$ GHz and $\nu = 1.4$ GHz at a
resolution of 4\arcsec.5 $\times$ 4\arcsec.5. The lighter the
grey, the steeper the spectral index. We excluded the region in
which the error is $>0.2$. Superposed are the contours of the 8.4
GHz total intensity (with the same levels as in Fig.
\ref{mappa-8.4-D}). } \label{index}
\end{figure}

In Fig. \ref{index} we show a greyscale of the spectral index map
between 8.4 and 1.4 GHz at $\sim$4.5\arcsec resolution
\footnote{The imaging procedure at each frequency (8.4 GHz in
array A and 1.4 GHz in array A+B combined) was performed using
data with matched uv coverage} superposed on the 8.4 GHz total
intensity contours. Besides confirming the steep spectrum ($\alpha
\sim -1$) of the central region, this figure shows the general
trend of the spectrum of the diffuse emission to steepen toward
the outer region ($\alpha \simeq -1$ to $\simeq -1.5$). We
estimated that the very low brightness emission extended to the
north at 1.4 GHz in array B (Fig. \ref{zoom}a) has a steep
spectral index $\alpha$ \ltsim $-1.5$, by assuming an upper limit
of 3$\sigma$ for the 8.4 GHz flux.
\begin{table*}
\begin{center}
\caption{Radio results for RBS797}
\begin{tabular}{ccccccccc}
\hline
\hline
Freq. & Array & Beam & Size & rms & Peak & Total flux & Radio Power & Flux central emission \\
(GHz) &~& (\arcsec)  & (\arcsec$^2$) & (mJy/beam) & (mJy/beam) & (mJy)& ($10^{24}$W Hz$^{-1}$) &(mJy)\\
\hline
~&~&~&~&~&~&~&~&~\\
8.4 & D & $ 6.31 \times 3.79 $  & $28 \times 25$ & 0.009 & 1.53 &$3.02 \pm 0.03$ & $1.24 \pm 0.01$ & $1.44 \pm 0.02$ \\
1.4 & B & $ 4.94 \times 3.81 $  & $44 \times 30$ & 0.027 & 9.94 &$26.04 \pm 0.26$ & $10.7 \pm 0.11$ & $11.49 \pm 0.12$  \\
1.4 & A & $ 1.52 \times 1.14 $  & $18 \times 11$ & 0.017 & 4.83 &$17.90 \pm 0.18$ & $7.38 \pm 0.07$ & $10.75 \pm 0.11$ \\
4.8 & A & $ 0.39 \times 0.35 $ & $5.5 \times 1.5$ & 0.013 & 0.98 &$2.63 \pm 0.03$& $1.08 \pm 0.01$  & $0.83 \pm 0.02$ \\
\hline
\label{risradio.tab}
\end{tabular}
\end{center}
\end{table*}

%%%%%%%%%%%%%%%%%%%%%%%%%%%%%%%%%%%%%%%%%%%%%%%%%%%%%%%%%%%%%%%%%%%%%%%%%%%%%%

\section{Discussion}
\label{discussion}

%%%%%%%%%%%%%%%%%%%%%%%%%%%%%%%%%%%%%%%%%%%%%%%%%%%%%%%%%%%%%%%%%%%%%%%%%%%%%

\subsection{X-rays}
\label{xray}

The cluster RBS797 was observed in 2000 with the \textit{Chandra}
Advanced CCD Imaging Spectrometer (ACIS) I detector (Schindler et
al. 2001). The smoothed X-ray image (0.5-7 keV) produced from
these data is shown in Fig. \ref{overlay}.
At large radii, RBS797 shows a very regular morphology, looking
fairly relaxed with a slight elliptical shape. The inner part of
the cluster shows the high surface brightness characteristic of a
cooling flow, even though there might be contamination due to the
emission of a central AGN (Schindler et al. 2001). At small radii,
the X-ray emission shows a peculiar structure instead, with two
strong depressions in the NE-SW direction at distances of about
3-5 arcsec from the cluster center. The minima are opposite to
each other with respect to the cluster center. In perpendicular
directions, a bar of enhanced X-ray emission is visible: the
surface brightness at radii of 4 arcsec is a factor of 3 - 4 lower
in the minima compared to the perpendicular directions, which
corresponds to about 20 counts in each minimum and 70 counts in
the same area in perpendicular directions (Schindler et al. 2001).
The X-ray depressions, which have a size of about 15-25
kpc\footnote{Quantities were scaled accordingly to the value of
$H_0$ adopted.}, are surrounded by a bright ring of emission. The
brightest parts of the ring and bar appear to form two shells to
the NE and SW.

A recent re-analysis of these data was performed by B\^{\i}rzan et
al. (2004), who find a cooling radius $r_{\rm cool} \sim 190$ kpc.
By making a deprojected spectral analysis, they estimated a
spectroscopic cooling luminosity $L_{\rm spec} \sim 1.2 \times
10^{45} {\rm erg \, s}^{-1}$ and a total luminosity inside the
cooling radius $L_{\rm X} \sim 4.5 \times 10^{45} {\rm erg \,
s}^{-1}$.
A visual inspection of the X-ray image led them to infer a size of
$\sim 10 \times 10$ kpc and $\sim 14 \times 9$ kpc for the two
cavities, located at a projected distance from the radio core of
$\sim 20$ kpc and $\sim 24$ kpc, respectively. The ratio of the $p
V$ work done on the surrounding medium by the cavity (calculated
from measurements of the cavity volume and gas pressure) to the
age of the cavity gives the instantaneous mechanical luminosity.
By assuming for the age the time required for the cavity to rise
buoyantly at its terminal velocity, B\^{\i}rzan et al. (2004)
estimate a mechanical luminosity for the cavity pair observed in
RBS797 of $L_{\rm mech} \sim 2.8 \times 10^{44} {\rm erg \,
s}^{-1}$.

%%%%%%%%%%%%%%%%%%%%%%%%%%%%%%%%%%%%%%%%%%%%%%%%%%%%%%%%%%%%%%%%%%%%%%%%%%%%%%

\subsection{Radio}
\label{radio}

The multifrequency VLA observations of RBS797 presented in Sect.
\ref{results} clearly reveal the presence of radio emission on
three different scales.
The lowest resolution images show diffuse radio emission extended
on a scale of hundreds of kpc (see Figs. \ref{mappa-8.4-D} and
\ref{zoom}a). The size of the extended radio emission, which is
characterized by amorphous morphology and a steep spectrum that
steepens with distance from the center, is roughly comparable to
that of the cooling region (B\^{\i}rzan et al 2004). These
characteristics point to a possible classification of the diffuse
radio source as a mini-halo. In addition to radio halos and
relics, radio mini-halos represent another class of diffuse radio
sources associated with the ICM. They have low surface brightness,
steep spectrum, and they are extended on a scale up to $\sim 500$
kpc, surrounding a dominant radio galaxy at the cluster center.
They are only observed in clusters with a cooling flow.

Besides the large-scale emission, radio emission extended on a
scale of tens of kpc is detected at arcsec resolution (Fig.
\ref{zoom}b). In particular, there is clear evidence of the
presence of a bright inner radio emission that is surrounded by a
more extended, fainter emission elongated in the NE-SW direction.
Observations at even higher (subarcsec) resolution allow us to
clearly distinguish the innermost radio jets on the kpc scale
(Fig. \ref{zoom}c) that point to the N-S direction.

%%%%%%%%%%%%%%%%%%%%%%%%%%%%%%%%%%%%%%%%%%%%%%%%%%%%%%%%%%%%%%%%%%%%%%%%%%%%%%

\subsection{X-ray -- radio interaction}
\label{interaction}

\begin{figure}[ht]
\includegraphics{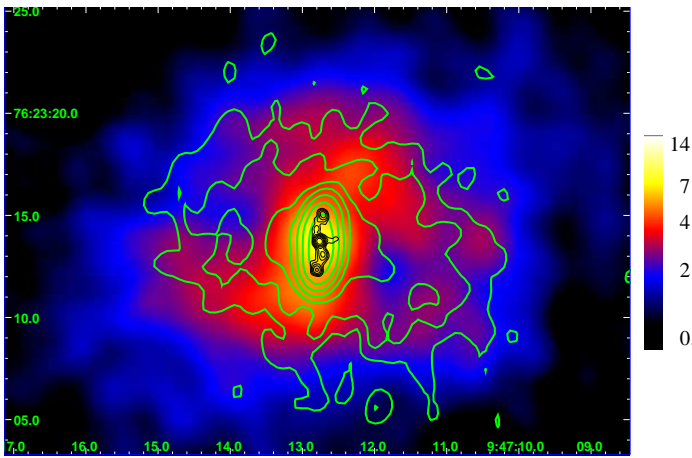}
\vspace{6.8cm}
\caption{
1.4 GHz (green) and 4.8 GHz (black) radio contours overlaid onto the
\textit{Chandra} X-ray image (0.5-7 keV) of the central
region of RBS797.
The X-ray surface brightness image, in units of cts/s/arcmin$^2$,
is smoothed with a Gaussian of $\sigma = $0.75 arcsec.}
\label{overlay}
\end{figure}

The new radio images presented here for RBS797 led us to classify
the cavities observed in X-rays as radio-filled cavities, contrary
to their classification as ghosts in the sample by B\^{\i}rzan et
al. (2004).
The classification as radio-filled cavities appears evident in
Fig. \ref{overlay}, where the 1.4 GHz radio contours are overlaid
onto the \textit{Chandra} X-ray image of the inner region of
RBS797.

It is very interesting to note that the radio structure is
elongated exactly towards the holes detected in the X-ray
emission; in particular, most of the extended radio emission is
projected within the X-ray holes to the NE-SW. Almost all of the
radio emission is contained within the bright X-ray shells.
Besides the 1.4 GHz radio contours at arcsec resolution, in Fig.
\ref{overlay} we also show the 4.8 GHz radio contours at subarcsec
resolution overlaid onto the \textit{Chandra} X-ray image. This
radio--X overlay is indicative of a strong interaction between the
ICM and the radio source embedded in the central cluster galaxy.
In particular, it suggests that the expansion of the 1.4 GHz radio
source has displaced the thermal gas that was formerly in the
X-ray holes and compressed this gas into the bright shells.
On the other hand, it is very remarkable that the two jets visible
at 4.8 GHz are oriented in a direction perpendicular to the
cavities, thus indicating that either the jet axes have been
deflected from their original directions into the outer lobes or
the renewed activity has begun at a different position angle from
the original outburst that created the outer radio lobes.
If confirmed, the small 4.8 GHz structure visible at high
resolution extending to the west of the nucleus perpendicular to
the main jet axis  (see Fig. \ref{zoom}c) could be related to the
one observed on a larger scale at 1.4 GHz (see Fig. \ref{zoom}b).

The structure extending in the NE-SW direction, visible at 1.4 GHz
in Fig. \ref{overlay} may originate by the deflection of the inner
jets visible at 4.8 GHz (Fig. \ref{overlay}) oriented in the N-S
direction. A deflected jet scenario has been investigated in the
case of the radio source PKS 2322$-$123 located at the center of
the cooling flow cluster A2597 (Pollack et al. 2005).

An alternative theory to that of a deflected jet is a precessing
radio jet, as described by Gower et al. (1982). In this scenario,
the current stage of RBS797 could lead to the creation of future
``ghost cavities'', which are thought to be buoyant lobes from a
past outburst of the radio galaxy and which have displaced the
thermal gas as they rose through the cluster atmosphere.
Observational evidence suggests that radio galaxies undergo
episodic outbursts that last $\sim 10^7$ yr and also have
repetition intervals of $\sim 10^8$ yr (e.g., McNamara et al.
2001). Over the cluster lifetime, the activity of the central
radio galaxy would produce many generations of radio lobes. This
is consistent with the observation both of the random location of
older, more distant bubbles and the irregular X-ray images within
the central few kpc of some galaxy clusters (e.g. Perseus:
Churazov et al. 2000, B\"ohringer et al. 1993, Fabian et al.
2000), which are apparently inconsistent with nonthermal jets
having fixed orientations defined by the spin axis of massive
black holes (Brighenti \& Mathews 2002).

Precessing inner jets could be explained in the context of binary
black hole models. The presence of a binary black hole in a
galactic nucleus may indeed become manifest through precession of
a jet propagation along the rotation axis of the primary, more
massive black hole during a period of activity (Begelman,
Blandford \& Rees 1980). Hence, relativistic material ejected in
different episodes of outbursts may be expelled in different
directions from the central engine similar to the radio directions
in the observations here. The picture of RBS797 harboring a
supermassive binary black hole might be further strengthened by
the small feature extending to the west in Fig. \ref{zoom}a, which
could itself be interpreted as ejecta originating from the
secondary black hole.

As discussed above, the different radio jet orientation observed
in RBS797 could be explained either by deflection or precession.
For both these cases, one important question is how violent and
energetic the expansion of the radio source after the outburst is.
Heinz, Reynolds, \& Begelman (1998) and Rizza et al. (2000)
suggested that the radio source would create cavities in the ICM
by highly supersonic expansion into the gas and that such a
violent expansion would drive strong shocks into the ICM. Due to
the compression in the shocks, the shell of X-ray gas surrounding
the radio-filled cavity would have a higher temperature, pressure,
and specific entropy than would the gas just outside the shell.
This model seems to be ruled out in the case of the 1.4 GHz radio
source in RBS797, as Schindler et al. (2001) find a slightly lower
temperature in the bright shells than for the rest of the cluster.
Since the gas in the shells is denser than the ICM outside the
shells, the specific entropy ($S \propto {\rm T} {\rm n}^{-2/3}$)
is actually much lower than outside the shells, suggesting that
the shells are probably not shocked regions. Thus, the expansion
of the radio source is probably subsonic or mildly transonic, as
found in other clusters, such as Hydra A (McNamara et al. 2000,
David et al. 2001), Perseus (Fabian et al. 2000), and A2052
(Blanton et al. 2001).
However, due to the short exposure time of the \textit{Chandra}
observation, it is not possible to derive a temperature map, which
would be essential for distinguishing different models.
Furthermore, one might expect the X-ray gas to have equal or
greater pressure than the non-thermal gas, which would be
consistent with the hypothesis of a confined radio jet.
Unfortunately with the X-ray data currently available, it is not
possible to perform an accurate study of the pressure balance
between the X-ray emitting gas and the innermost radio source in
RBS797.

One possible qualitative interpretation we can infer from the
present data is that the 1.4 GHz emission extending to the NE-SW
from the center consists of bubbles of radio-emitting plasma
created by twin jets in the past, which are moving away from the
center due to buoyancy, while the two jets visible at 4.8 GHz are
related to current nuclear activity. The qualitative picture is
that a past radio activity may have injected radio-emitting plasma
that then propagates through the cooling flow region in the form
of buoyant subsonic plumes. During the initial phase, the jets
inflate the cocoon with relativistic plasma, and the expansion is
supersonic, while at a later stage the expansion slows down and
becomes subsonic. On the other hand, buoyancy limits the growth of
the cavities inflated by the jets, and when the rising velocity
due to buoyancy exceeds the expansion velocity, the bubble
detaches from the jet and begins rising (Churazov et al. 2000). As
shown by several authors who have studied the hydrodynamic
evolution of rising bubbles (e.g. Gull \& Northover 1973; Churazov
et al. 2000; Br\"uggen \& Kaiser 2001), the moving away from the
center is accompanied by adiabatic expansion and further mixing of
the energetic radio source material with the ambient ICM. This
latter process occurs via Kelvin-Helmholtz (K-H) instabilities on
the surface of the bubbles and, more importantly, via
Rayleigh-Taylor (R-T) instabilities on the "top" surface of the
bubble that will actually disrupt and fragment it. Hydrodynamic
simulations show a dramatic breakup and fine-scale mixing of the
bubble material with the ICM in about $10^7$ - $10^8$ years (e.g.
Br\"{u}ggen \& Kaiser 2002). This is comparable to (or bigger
than) the lifetime of the electrons that produce synchrotron
radiation.\footnote{ The radiative lifetime of an ensemble of
relativistic electrons losing energy by synchrotron emission and
inverse Compton (IC) scattering off the CMB photons, defined such
as $\tau = - E/\dot{E}$, is given by (e.g., Ginzburg \&
Syrovatskii 1965, Rybicki \& Lightman 1979):
$$
\tau_{\rm sin+IC} \simeq \frac{2.5 \times 10^{13}}{\left[\left(
B/\mu\mbox{G}\right)^2
+ \left(B_{\rm \tiny CMB}/\mu\mbox{G}\right)^2 \right] \, \gamma}
 \; \; \; \mbox{yr}
$$
where $B$ is the magnetic field intensity, $\gamma$ the Lorentz
factor, and $B_{\rm \tiny CMB} = 3.18 (1+z)^2 \mu$G the magnetic
field equivalent to the CMB in terms of electron radiative
losses.} Electrons emitting at frequencies $\sim$ 1 GHz in
magnetic fields on the order of $\sim$ 1-3 $\mu$G, typical for the
ICM, have a radiative lifetime of $\sim 7 \times 10^7$ yr.

The disruption of the bubbles produced in past radio outbursts
would then have left a population of relic relativistic electrons
mixed with the thermal plasma in the cooling flow region. These
relic electrons could diffuse in the thermal plasma up to $\sim
100$ kpc scale in a few Gyr (e.g., Brunetti 2003) filling the
whole region observed in Fig. \ref{zoom}a and thus forming the
population of relic electrons required by the re-acceleration
model for the origin of radio mini-halos\footnote{Alternatively,
Pfrommer \& En{\ss}lin (2004) suggest that radio mini-halos may be
of hadronic origin from cosmic rays.}
 (Gitti et al. 2002, 2004).
In particular, analogous to the case of the Perseus cluster, it
might be difficult to find a direct connection between the inner
radio lobes and the large-scale extended emission in terms of
simple buoyancy or particle diffusion, since the expansion and
buoyancy of blobs would produce adiabatic losses and a decrease in
the magnetic field causing too strong a steepening of the
spectrum, which would in turn prevent detection of large-scale
radio emission. Thus, if the relativistic electrons are of primary
origin, efficient re-acceleration mechanisms in the cooling flow
region are necessary to explain the presence of the large-scale
radio emission. A detailed application to RBS797 of the model for
the origin of radio mini-halos consisting of MHD turbulent
re-acceleration of intra-cluster cosmic ray electrons (Gitti et
al. 2002, 2004)  will be presented in a forthcoming paper.

%%%%%%%%%%%%%%%%%%%%%%%%%%%%%%%%%%%%%%%%%%%%%%%%%%%%%%%%%%%%%%%%%%%%%%%%%%%%%%

\subsection{Can radio source heating quench the cooling?}

By assuming a spectral index $\alpha = -1$, we calculated the
total radio luminosity of the full source visible in Fig.
\ref{zoom}a over the frequency range of 10 MHz - 10 GHz to be
$L_{\rm radio} \sim 10^{42} {\rm erg ~ s}^{-1} $. In order to
investigate whether the radio source deposits enough energy into
the ICM to quench cooling, this has to be compared with the
cooling luminosity. As derived by B\^{\i}rzan et al. (2004) in the
case of RBS797, the cooling luminosity is $L_{\rm X} - L_{\rm
spec} \sim 3.3 \times 10^{45} {\rm erg \, s}^{-1}$, therefore the
radio source is too weak by a factor of a few thousand to balance
cooling in this system.

On the other hand, it is well known that radio sources are
inefficient radiators and that the mechanical (kinetic) luminosity
$L_{\rm mech}$ of radio sources can be much higher than their
synchrotron luminosity. One way to estimate $L_{\rm mech}$ is
through measurements of the X-ray cavity size and surrounding gas
pressure, as done by B\^{\i}rzan et al (2004) for a sample of
cavity clusters. In particular, these authors derive a value of
$L_{\rm mech} \sim 2.8 \times 10^{44} {\rm erg \, s}^{-1}$ for the
cavity pair observed in RBS797. We estimated the radio luminosity
corresponding to the radio structure visible at 1.4 GHz in array A
(Fig. \ref{overlay}) to be $\sim 7.1 \times 10^{41} {\rm erg ~
s}^{-1} $. The mechanical luminosity $L_{\rm mech}$ is thus $\sim
400$ times the observed radio luminosity of the radio source
filling the X-ray cavities. Even considering the total mechanical
luminosity of the central radio source, as estimated from the
cavity properties, it seems that cooling cannot be balanced by
bubble heating in the galaxy cluster RBS797.

%%%%%%%%%%%%%%%%%%%%%%%%%%%%%%%%%%%%%%%%%%%%%%%%%%%%%%%%%%%%%%%%%%%%%%%%%%%%%%

\section{Conclusions}

We have presented multifrequency (1.4, 4.8, and 8.4 GHz) VLA
observations of the radio source located at the center of the
cooling flow cluster RBS797 (z=0.35), which is the first distant
cluster in which two pronounced X-ray depressions have been
discovered by previous \textit{Chandra} observations (Schindler et
al. 2001). We detected radio emission on three different scales
showing orientation in different directions, thus indicating that
RBS797 represents a very peculiar case.

The highest resolution image on kpc scale shows the details of the
innermost 4.8 GHz radio jets, which point clearly to the N-S
direction. Remarkably, the inner jets are oriented in a direction
almost perpendicular to that of the extended structure detected at
lower resolution, which is elongated in the NE-SW direction
exactly towards the holes detected in X-rays. The lowest
resolution images show large-scale radio emission with amorphous
morphology, slightly extended in the N-S direction. We therefore
find evidence of a strong interaction between the central radio
source and the intra-cluster medium in RBS797.

Further observations in X-rays and radio are needed to either rule
out or confirm the models that we have suggested for the
interaction between the radio source and the X-ray gas. In
particular, more detailed investigations of the innermost
structure of this source will add an important piece of
information for investigating the energetics involved in the
formation of cavities in the X-ray gas and in the process of
interaction between the radio sources and the ambient gas.

%%%%%%%%%%%%%%%%%%%%%%%%%%%%%%%%%%%%%%%%%%%%%%%%%%%%%%%%%%%%%%%%%%%%%%%%%%%%%%
%%%%%%%%%%%%%%%%%%%%%%%%%%%%%%%%%%%%%%%%%%%%%%%%%%%%%%%%%%%%%%%%%%%%%%%%%%%%%%

\begin{acknowledgements}
We thank W. Domainko for suggesting the possibility of RBS797 harboring a
binary black hole, and G. Hasinger for useful discussions.
We also thank the anonymous referee for useful comments.
M.G. would like to thank F. Brighenti
for many stimulating discussions and insightful comments.
This work was supported by the Austrian Science Foundation FWF under
grant P15868 and by the Tiroler Wissenschaftsfonds.
\end{acknowledgements}

%%%%%%%%%%%%%%%%%%%%%%%%%%%%%%%%%%%%%%%%%%%%%%%%%%%%%%%%%%%%%%%%%%%%%%%%%%%%%%

\end{document}